# Identifying modulation formats using integrated clustering algorithm

WENBO ZHANG,[1,3,*] JINMEI YE,[1,3] ZIXIAN YUE,[1,3] YUXIANG WANG,[1,3] XULUN ZHANG,[2,3] XIAOGUANG ZHANG,[2,3] LIXIA XI,[2,3]

[1]*School of Science, Beijing University of Posts and Telecommunications, China*

[2] *School of Electronic Engineering, Beijing University of Posts and Telecommunications, China*

[3]*State Key Laboratory of Information Photonics and Optical Communications, Beijing University of Posts and Telecommunications, China*

*\*zhangwb@bupt.edu.cn*

Abstract: Modulation format identification (MFI) is crucial in next-generation optical networks such as cognitive optical networks. An integrated-clustering-algorithm-based MFI scheme in a coherent optical communication system is proposed herein. Numerical simulations are performed to test the performance of the scheme on the platform of a polarization domain multiplexing system at a symbol rate of 28 Gbaud. Simulations show that the MFI scheme can achieve an accuracy of 100% for five modulation formats considered in this study when the optical signal-to-noise ratio values are lower than the 7% forward error correction threshold. Tolerances to residual chromatic dispersion (CD) are discussed, and the results show that the proposed scheme can perform well as the residual CD changes over a wide range of values. Moreover, by selecting a number of key blocks, the proposed scheme is less complex than other clustering-based MFI schemes.

1. Introduction

Driven by new applications such as cloud computing, internet of things, and self-driving, the demand for data transmission has increased rapidly. Dynamic networks, particularly the dynamic optical network (DON), which provides high capacity and efficiency, have become indispensable. To satisfy the requirements of the DON, an intelligent cognitive optical network (CON) with autonomous sensing and control has been proposed [1–2]. By understanding the missions and environments, the configuration of the network can be changed automatically. For instance, a transmitter exhibits flexibility in terms of signal bandwidth, data rate, modulation format, transmission wavelength, signal power, etc. The key to implementing CONs is to design a flexible transmitter and receiver [1–2]. When the characteristics of the network change, the receiver must change its configuration automatically because a large number of digital signal processing (DSP) algorithms are designed based on these network characteristics. For example, the modulation format (MF) is one of the most important characteristics. To automatically adjust the configuration at the receiver, a data-driven modulation format identification (MFI) scheme that only depends on the received signals is urgently required.

MFI has been investigated widely. In [3], a pilot-aided MFI scheme is proposed. In [4, 5], an MFI scheme using higher-order cumulants of the received signals is proposed. In [6–11], some MFI schemes based on mapping received signals into a three-dimensional Stokes space are presented. In addition, some supervised machine learning (ML) schemes using neural networks [11–13] and support vector machines [14] have been proposed. However, the critical problem in supervised ML schemes is that they must be trained by using a large number of labeled data before they are implemented. To overcome this disadvantage, some unsupervised clustering algorithms [15–18] have been used to design the MFI scheme.

Herein, an unsupervised clustering algorithm-based MFI scheme is proposed. By selecting a number of key blocks of received signals on a constellation plane, the scheme becomes less complex than other clustering-based schemes. Simulations show that the proposed scheme can achieve an accuracy of 100% for each modulation format since the optical signal-to-noise ratio (OSNR) is lower than that in [17]. The tolerances of the proposed scheme to impairments of residual CD are also discussed, and the results show that the proposed scheme can achieve 100% identification accuracy as the value of residual CD changes in the range from -400 to 350 ps/nm.

In Section 2, the basic principle and architecture of the MFI scheme are introduced. In Section 3, the details of the simulation platform are presented. In Section 4, the performance of the proposed scheme is discussed, and conclusions are provided in Section 5.

2. Principle and MFI scheme

*2.1 Two-dimensional (2D) histogram of received signals and key blocks*

Each modulation format has its own distribution character in the constellation plane and can be used to distinguish it from others. These characteristics can be visualized clearly using a 2D histogram. In Fig. 1, the constellation plane is separated into an $80 \times 80$ grid of equally spaced blocks, and the modulation formats can be determined by counting the number of bin clusters. As shown in Fig. 1, the number of clusters for QPSK, 8PSK, 16QAM, 32QAM, and 64QAM are 4, 8, 16, 32, and 64, respectively. However, these conclusions may not be clear in cases where the OSNR is low because noise blurs the constellations.

Fig. 1 2D histogram of received signals of (a) QPSK, (b) 8PSK, (c) 16QAM, (d) 32QAM, and (e) 64QAM, where OSNR is 30 dB.

To reduce the effects of impairments, 640 bins with significant heights in the 2D histogram were used as key bins. As shown in Fig. 1(a), the 2D histogram contains four peaks, and all bins with significant heights surround these peaks. Bins with significant heights are known as key bins. Key bins can be obtained by sorting all bins with respect to their heights in the descending order and selecting the first 640 bins. The location of each key bin is known as a key block. In Fig. 2, the key blocks for QPSK, 8PSK, 16QAM, 32QAM, and 64QAM are indicated as red spots, and the OSNR was set to 20 dB. As shown in Figs. 2(a) to (d), key blocks accumulated into 4, 8, 16, and 32 clusters, respectively. The number of clusters of key blocks in Fig. 2(e) may not be exactly 64 because the OSNR is extremely low for 64QAM.

Fig. 2 Key blocks in constellation plane for QPSK, 8PSK 16QAM, 32QAM, and 64QAM, where OSNR was 20 dB.

*2.2 MFI scheme*

The proposed MFI scheme is designed for the PDM *m*-PSK and PDM *m*-QAM coherent systems, as shown in Fig. 3. After CD compensation, Pol-deMUX, the proposed MFI scheme was used to extract information regarding the modulation format. After implementing the MFI scheme, other DSP processes can be implemented based on the modulation format.

Fig. 3 DSP model of proposed MFI scheme.

As shown in Fig. 2, the key blocks of the received signals can be categorized into several clusters. Although key blocks of 16QAM (Fig. 2(c)) can be classified into any number of clusters, 16 might be the optimal number. Similarly, 4 is the optimal number of clusters for QPSK, 8 for 8PSK, etc. Based on these results, the proposed MFI scheme determines the optimal number of key block clusters; subsequently, this number is used as a character to identify the modulation format.

Fig. 4 Architecture of proposed MFI scheme.

The architecture of the proposed MFI scheme (see Fig. 4) can be implemented in two stages. In the first stage, key bins and key blocks are selected from the 2D histogram of the received signals. In the second stage, the MFI result is provided by selecting the optimal number of key block clusters.

In the first stage, two steps are involved. In the first step, 4QAM-BPS acts on the received signals. The effects of the 4QAM-BPS are shown in Section 2.3. In the second step, a 2D histogram of the results of the first step is generated, and key bins are selected from the histogram. A set of key blocks is generated by selecting the locations of key bins.

In the second stage, the set of key blocks is accumulated into $k$ clusters using the nearest neighbor prototype rule, where $k$ is a positive integer. The result is known as a $k$-partition and is denoted by $P_k$. As $k$ increases with each step, a series of $k$-partitions can be obtained, where $k = 1, ..., m$, and the maximum number of partitions $m$ is 100. To evaluate the performance of each $k$-partition, an evaluation function $f$ is defined, as detailed in Section 2.4. The value of $k$ that yields a maximum $f(P_k)$ value is known as the best $k$ and is denoted by $k^*$. The process to determine the best $k$ is illustrated in Algorithm 1.

## 2.3 Effects of 4QAM-BPS

Because no information regarding the MF is available prior to the MFI scheme, to improve the tolerance of the MFI scheme with respect to impairments caused by the phase noise of the laser, the 4QAM-BPS was used in the first stage. The received signals of QPSK, 8PSK, 16QAM, 32QAM, and 64QAM and their corresponding results after the 4QAM-BPS are shown in Fig. 5. As shown, the constellation points accumulated after the 4QAM-BPS. Therefore, the 4QAM-BPS can improve the performance of the proposed MFI scheme.

## 2.4 Evaluation function of partitions

To evaluate the performance of a partition, several indexes are typically used [19]. In this study, we used the Silhouette [20] index.

Let $P$ be a partition of dataset $D$, and $C = \{c_1, c_2, ..., c_{|C|}\}$ be the set of cluster centers of $P$, where $|C|$ is the number of cluster centers. The clustering algorithm assigns each element to

Fig. 5 Constellations after modulation-independent phase processing.

$D$ to its closest cluster center $c_p \in C$. Therefore, if $s_i \in D$ and it is associated with the cluster center $c_p$, then $\|s_i - c_p\| \leq \|s_i - c_q\|$, $\forall p, q = 1, 2, ..., |C|$, $p \neq q$, where $\|s_i - c_p\|$ is the Euclidean distance between $s_i$ and $c_p$. The notation $s_i \mapsto c_p$ indicates that element $s_i$ is associated with the cluster center $c_p$, and $|c_p|$ is the number of all elements associated with $c_p$. For each $c_p$, how close an element $s_i$ associated with $c_p$ matches with other elements related to $c_p$ can be measured by the average within-cluster dissimilarity (cohesion).

$$In_i = \frac{1}{|c_p|} \sum_{\substack{s_i \mapsto c_p \\ s_j \mapsto c_p}} \|s_i - s_j\|.$$

How well an element $s_i$ is associated with its closest neighboring clusters $c_q$ can be measured by the smallest average between-cluster dissimilarity (separation).

$$Out_i = \min \left\{ \frac{1}{|c_q|} \sum_{\substack{s_i \mapsto c_p \\ s_j \mapsto c_q \\ p \neq q}} \|s_i - s_j\| \right\}.$$

For each $s_i$, the Silhouette index is defined by combining the cohesion and separation as follows:

$$Silh_{s_i} = \frac{Out_i - In_i}{\max(In_i, Out_i)}.$$

This value is in the range $[-1,1]$, and the larger the Silhouette index, the better $s_i$ fits within its own cluster. Finally, the evaluation function of the partition $P$ is defined as

$$f(P) = \frac{1}{|D|} \sum_{s_i \in D} Silh_{s_i} = \frac{1}{|D|} \sum_{s_i \in D} \frac{Out_i - In_i}{\max(In_i, Out_i)},$$

where |D| is the number of elements in D, i.e., the average of the Silhouette index over D.

3. Setup of simulation platform

The simulation platform setup is shown in Fig. 6. The 28 Gbaud signals of five modulation formats, including PDM-mPSK signals (QPSK and 8PSK) and PDM-mQAM signals (16QAM, 32QAM, and 64QAM), were generated using a 65 GS/s arbitrary wave generator. At the transmitter, external cavity lasers (ECLs) were utilized to produce light at a central wavelength of 1550 nm and then modulated by an integrated polarization-multiplexing I/Q modulator. The linewidth of the ECLs was 100 kHz. The transmitted signals were amplified by an EDFA and sent into a standard single-mode fiber (SSMF). The attenuation coefficient and CD parameter of the fiber were 0.2 dB/km and 16.9 ps/(nm·km), respectively. At the receiver side, a bandwidth optical bandpass filter of approximately 33 GHz was used to reject the out-of-band ASE noise, and the filtered signals were fed into a 42 GHz electrical bandwidth polarization diversity coherent receiver. A local oscillator with a bandwidth of 100 kHz was used. After an 80 GS/s of analog-to-digital sampling was performed using a real-time oscilloscope, electrical signals were processed by an offline DSP module, where the proposed MFI was embedded. Inside the DSP module, CD compensation and Pol-deMUX [21] were performed first, and the 4QAM-BPS was used for processing the phase noise in advance. Subsequently, the proposed MFI scheme was used to extract the modulation format and other DSP processes based on the modulation format.

Fig. 6 Setup of simulation platform.

4. Performances of MFI scheme

*4.1 Thresholds for MFI*

In the second stage of the proposed MFI scheme, the output depends on the value of $k^*$ provided by the first stage. In Fig. 7, the results of $k^*$ of five modulation formats are shown, where the OSNR is in the range from 4 to 30 dB. For example, when the OSNR is 19 dB, the $k^*$ obtained by the proposed scheme for QPSK, 8PSK, 16QAM, 32QAM, and 64QAM modulation formats are 4, 8, 16, 31, and 4, respectively. Therefore, if the value of $k^*$ is 8 or 16, then the modulation format can be determined as 8PSK or 16QAM, respectively. If $k^*$ is 31, then the modulation format must be 32QAM; if $k^*$ is 4, the modulation format may be QPSK or 64QAM. To determine the output of the second stage, the decision regions with respect to the value of $k^*$ were listed, as shown in Table 1.

Fig. 7 Values of $k^*$, where modulation formats are QPSK, 8PSK, 16QAM, 32QAM, and 64QAM; OSNR ranged from 4 to 30 dB.

Table 1 Thresholds for MFI

*4.2 Performances of MFI*

The performance of the proposed MFI was verified in a 28 Gbaud optical transmission system (see Fig. 6). The OSNR ranges of the PDM-QPSK, PDM-8PSK, PDM-16QAM, PDM-32QAM, and PDM-64QAM signals were set from 4 to 30 dB. To verify the accuracy and reliability of the algorithm, we generated 100 datasets for each of the five modulation formats, i.e., $24 \times 100$ independent datasets for each modulation format. As shown in Fig. 8, the MFI scheme achieved an accuracy of 100% for all five modulation formats when the OSNRs were lower than the 7% FEC threshold.

Fig. 8 Identification accuracy for modulation formats of PDM-QPSK, PDM-8PSK, PDM-16QAM, PDM-32QAM, and PDM-64QAM signals at different OSNRs (4 to 30 dB).

In the following, the tolerance of the proposed MFI scheme with respect to the residual CD [22] is discussed. In Fig. 9, the identification accuracy vs. residual CD is shown. The OSNRs of five modulation formats, PDM-QPSK, PDM-8PSK, PDM-16QAM, PDM-32QAM, and PDM-64QAM were 13, 16, 19, 22, and 25 dB, respectively. The values of residual CD ranged from -1000 to 1000 ps/nm. The results show that the proposed MFI scheme can achieve 100% identification accuracy when the value of the residual CD ranged from -400 to 350 ps/nm.

Fig. 9 Identification accuracy with residual CD for modulation formats of PDM-QPSK, PDM-8PSK, PDM-16QAM, PDM-32QAM, and PDM-64QAM signals.

*4.3 Complexity analysis*

By comparing the runtime of algorithms, complexity of some typical clustering algorithms for MFI including *k*-means, expectation maximization, OPTICS, Density Based Spatial Clustering of Applications with Noise (DBSCAN), spectral clustering and maximum likelihood clustering were discussed in [16]. As in [16], the DBSCAN algorithm has the lowest complex in computing. Herein, we compare the relative complexity between proposed MFI scheme and DBSCAN algorithm. The results are shown in Fig. 10, where the relative runtime is obtained by calculating average runtime and normalizing with respect to the slowest time for five modulation formats. Proposed MFI scheme significantly reduces the computation by extracted key bins from received signal and results show that the proposed MFI scheme has a relatively low and stable complexity.

Fig.10 The complexity analysis of proposed MFI scheme and DBSCAN algorithm for five modulation formats.

5. Conclusion

Herein, a clustering-based MFI scheme was proposed. By constructing a 2D histogram of received signals in the constellation plane, the character of the modulation format can be captured by selecting a number of key bins. The modulation format was identified by the number of clustering centers of key bins, resulting in the maximum value of the evaluation function. Because the number of key bins was significantly smaller than that of the received signals, the proposed scheme was significantly less complex than the other clustering-method-based MFI schemes. Simulations showed that the proposed scheme achieved a 100% accuracy when the OSNR was lower than the 7% FEC threshold for each modulation format (i.e., PDM-QPSK, PDM-8PSK, PDM-16QAM, PDM-32QAM, and PDM-64QAM). In addition, the tolerances of the proposed scheme to the residual CD indicated the highly stable performance of the proposed scheme.